\documentclass[11pt, a4paper]{article}
\pdfoutput=1
\usepackage{jcappub}
\usepackage{amssymb}
\usepackage{amsmath}
\usepackage{braket}
\usepackage{listings}
\usepackage{cases}
\usepackage{comment}

\newcommand{\dif}[2]{\frac{\mathrm{d} #1}{\mathrm{d} #2}}
\newcommand{\pdif}[2]{\frac{\partial #1}{\partial #2}}
\newcommand{\dd}{\mathrm{d}}
\newcommand{\ee}{\mathrm{e}}

\newcommand{\ltsim}{\protect\raisebox{-0.5ex}{$\:\stackrel{\textstyle <}{\sim}\:$}}
\newcommand{\gtsim}{\protect\raisebox{-0.5ex}{$\:\stackrel{\textstyle >}{\sim}\:$}}

\title{Can massive primordial black holes be produced in mild waterfall hybrid inflation?}

\author{Masahiro Kawasaki}	
\author{and Yuichiro Tada}
\affiliation{Institute for Cosmic Ray Research, The University of Tokyo, Kashiwa, Chiba 277-8582, Japan}
\affiliation{Kavli Institute for the Physics and Mathematics of the Universe (WPI), UTIAS,
	The University of Tokyo, Kashiwa, Chiba 277-8583, Japan}
\emailAdd{kawasaki@icrr.u-tokyo.ac.jp}
\emailAdd{yuichiro.tada@ipmu.jp}

\abstract{
We studied the possibility whether the massive primordial black holes (PBHs) surviving today can be produced in hybrid inflation. 
Though it is of great interest since such PBHs can be the candidate for dark matter or seeds of the supermassive black holes in galaxies, 
there have not been quantitatively complete works yet because of the non-perturbative behavior around the critical point of hybrid inflation. 
Therefore, combining the stochastic and $\delta N$ formalism, we numerically calculated the curvature perturbations in a non-perturbative way 
and found, without any specific assumption of the types of hybrid inflation, PBHs are rather overproduced 
when the waterfall phase of hybrid inflation continues so long that the PBH scale is well enlarged 
and the corresponding PBH mass becomes sizable enough.
}

\arxivnumber{1512.03515}

\begin{document}

\begin{flushright}
IPMU 15-0205
\end{flushright}

\maketitle
\flushbottom

\section{Introduction}\label{Introduction}
The primordial black holes (PBHs) are theoretically suggested black holes produced in the early universe. 
They are formed by the gravitational collapse of the Hubble patches which are $\mathcal{O}(1)$ denser than their surroundings in the radiation
dominant~\cite{Hawking:1971ei,Carr:1974nx,Carr:1975qj}. The PBH abundance is connected to the properties of the primordial curvature perturbations, which determines how rare such quite overdense patches are.
As an interesting point of the PBH, it is one of the well-studied candidates for dark matter (DM). While all windows for PBHs to be a main
component of DM seem to be closed recently~\cite{Griest:2013esa}, there may be some loopholes. For example, the constraints for around 
$10^{23\text{--}24}\,\mathrm{g}\sim10^{-10}M_\odot$ from neutron stars are still under discussion (see e.g. \cite{Capela:2013yf,Pani:2014rca,Capela:2014qea,Defillon:2014wla} as related works).

Another important motivation of PBHs is to explain the seeds of supermassive black holes (SMBHs). Most galaxies including our Milky Way are thought to 
possess one or a few SMBHs whose masses reach to $10^{6\text{--}9.5}M_\odot$ in their centers~\cite{Kormendy:1995er}, 
and moreover, such massive black holes have been found even at high redshifts as $z\sim6\text{--}7$~\cite{Fan:2001ff,Mortlock:2011va}. 
While the astrophysical production mechanism of (especially high-$z$) SMBHs has still been unknown, 
the literature suggested massive PBHs ($\sim10^5M_\odot$) can be the seeds of SMBHs~\cite{Bean:2002kx}.
Therefore, whether sufficiently massive PBHs
can be produced (\emph{naturally} if possible) or not is an important subject.

While various mechanisms to produce massive PBHs have been proposed 
(\emph{double inflation:} \cite{Kawasaki:1997ju,Yokoyama:1998pt,Kawaguchi:2007fz,Frampton:2010sw,Kohri:2012yw}, 
\emph{running mass:} \cite{Kohri:2007qn,Drees:2011hb}, \emph{curvaton:} \cite{Kawasaki:2012wr,Bugaev:2012ai}, 
\emph{gauge field production:} \cite{Linde:2012bt,Bugaev:2013fya}),
we will focus on hybrid inflation in this paper.
Hybrid inflation, which was originally proposed by Linde~\cite{Linde:1993cn}, is a combined model of chaotic and hilltop inflation.
In this model, the inflationary universe is driven by the false vacuum energy of the so-called waterfall field, represented as $\psi$ here, which is 
stabilized by the coupling to the other scalar inflaton, denoted by $\phi$, at first. 
Then, when the inflaton's vev becomes small due to the potential of itself and
the coupling between $\phi$ and $\psi$ gets to unable to stabilize the waterfall field, inflation will terminate by the second order phase transition of $\psi$.
Hybrid inflation is an attractive model in the point that the initial condition problem is improved well in this model even though it is small field inflation (namely the scalar fields' vev does not exceed the Planck scale).

The mainstream of hybrid inflation is models where the inflaton's slow-roll phase (called \emph{valley phase} generically) 
continues more than 60 e-folds and
the waterfall transition ends instantaneously. Among this type, the original model where the inflaton's potential is given by simple mass term 
predicts blue-tilted curvature perturbations
which have been excluded now, but the supersymmetric flat inflaton whose potential can be raised up logarithmically 
due to the Coleman-Weinberg correction 
can give a red-tilted spectrum~\cite{Dvali:1994ms},
and moreover, it has been suggested that the additional linear potential from the soft supersymmetry (SUSY) breaking 
can realize $n_s\sim0.96$~\cite{Buchmuller:2000zm}, 
which is in the Planck's sweet spot~\cite{Ade:2015xua}.
Another direction of realizations of hybrid inflation is the long-waterfall models~\cite{Clesse:2010iz,Kodama:2011vs,Mulryne:2011ni,Clesse:2012dw,Clesse:2013jra}. 
In these models, $\psi$'s potential is so flat 
that the waterfall phase continues more than 60 e-folds like
hilltop inflation.

In this paper, we will concentrate on the intermediate case, namely the mild-waterfall models where the waterfall phase continues 
more than a few e-folds but less than 60 e-folds. 
An attractive point of the mild case is that very massive PBHs can be produced a lot. 
That is because the perturbations can grow much around the phase transition critical point
due to the flatness of the (especially $\psi$'s) potential and such perturbations will be inflated during the long-lasting waterfall phase.
Though the massive PBH production in mild-waterfall hybrid inflation has been discussed 
for a long time~\cite{GarciaBellido:1996qt,Lyth:2010zq,Bugaev:2011qt,Guth:2012we,Halpern:2014mca,Clesse:2015wea}, there is still not a quantitatively complete work because of its non-perturbative difficulty. 
Around the critical point, the field perturbations affect the background dynamics itself,
therefore intrinsically the scalar fields cannot be treated perturbatively during the phase transition (see \cite{Martin:2011ib,Levasseur:2013tja} for example). 

Recently, in refs.~\cite{Fujita:2013cna,Fujita:2014tja},\footnote{Recently Vennin and Starobinsky has verified this formalism and found some analytic expressions 
with use of techniques of stochastic calculus~\cite{Vennin:2015hra}.} we have proposed some non-perturbative algorithm to calculate the power spectrum of curvature perturbations 
in the stochastic formalism~\cite{Starobinsky:1986fx,Nambu:1987ef,Kandrup:1988sc,Nakao:1988yi,Mollerach:1990zf,Linde:1993xx,Starobinsky:1994bd}.
Especially in ref.~\cite{Fujita:2014tja}, the power spectrum in mild-waterfall hybrid inflation was calculated without perturbative expansions 
with respect to $\phi$ and $\psi$.
Following these works, we perform a wide parameter search in this paper with use of our algorithm and conclude that PBHs are rather 
overproduced in most of the mild-waterfall cases.
Also we show the parameter constraints from the PBH constraints.

The rest of the paper is organized as follows. In section~\ref{Aspects of hybrid inflation}, we introduce hybrid inflation and 
briefly review the analytic approximations for the scalar dynamics and curvature perturbations, following ref.~\cite{Clesse:2015wea}. 
In section~\ref{Parameter search}, we perform a rough parameter search and find that PBHs are overproduced in the mild-waterfall cases.
Also the parameter constraints including the non-Gaussian effects are obtained as a main result of the paper.
In section~\ref{Example power spectrum}, we exemplify the power spectra of the curvature perturbations, which indicate that the allowed PBH mass scales are indeed small, with use of our algorithm.
Finally section~\ref{Conclusions} is devoted to conclusions.

\section{Aspects of hybrid inflation}\label{Aspects of hybrid inflation}
In this section, we would like to introduce hybrid inflation and its various aspects.
Throughout this paper, we do not assume any specific UV-theoretical motivation like SUSY and simply refer to the models whose potential 
is given by the following form as \emph{hybrid inflation}.
\begin{align}
	V(\phi,\psi)=V(\phi)+\Lambda^4\left[\left(1-\frac{\psi^2}{M^2}\right)^2+2\frac{\phi^2\psi^2}{\phi_c^2M^2}\right].
\end{align}
Here $\phi$ and $\psi$ are two real scalar field which are called \emph{inflaton} and \emph{waterfall} fields respectively, while
$\Lambda,\,M$, and $\phi_c$ are dimensionful model parameters.\footnote{\label{No of waterfall}This potential has $Z_2$ symmetry as $\psi\longleftrightarrow-\psi$ which is broken
at the end of inflation, and therefore it actually causes domain walls and spoils the standard cosmology. 
To avoid domain walls, usually one more real scalar is added as the waterfall fields and $SO(2)$ symmetry is imposed on them.
It replaces domain walls with cosmic strings but they are harmless if the vev of the waterfall fields is sufficiently small.
Moreover any topological defect might be avoided with much more d.o.f. for the waterfall direction.
In any case, these modifications of the potential
will cause only factor differences in respect of the inflationary phenomenology and we will consider this single waterfall field case for simplicity 
(the authors of \cite{Halpern:2014mca} concluded that the power spectrum is suppressed by factor $\mathcal{N}$ for the $\mathcal{N}$-waterfall field case).}
As the second term indicates, the sign of the $\psi$'s effective mass squared, $m_{\psi,\mathrm{eff}}^2|_{\psi\sim0}=\partial_\psi^2V|_{\psi\sim0}=2\frac{\Lambda^4}{M^2}
\left(\frac{\phi^2}{\phi_c^2}-1\right)$, is determined by $\phi$'s vev.
Namely, if $\phi$'s vev is larger than the critical value $\phi_c$, $\psi$ is stabilized to the origin due to its positive mass squared.
Then, with the pseudo-flat $V(\phi)$ which is minimized around the origin, the inflaton can undergoes a slow roll to its minimum and \emph{switch on}
the tachyonic property of the waterfall field when $\phi$ reaches $\phi_c$. Subsequently inflation will be ended by the second order phase transition
of $\psi$.  

The stage before $\phi$ reaches $\phi_c$ is called \emph{valley phase} and that after $\phi_c$ is referred as \emph{waterfall phase} generically.
Though the waterfall phase basically ends instantaneously due to the tachonic instability, the literatures~\cite{Clesse:2010iz,Kodama:2011vs,Mulryne:2011ni,Clesse:2012dw,Clesse:2013jra}
suggested the possibility of the long-lasting waterfall. In this paper, we concentrate on the mild-waterfall case where the waterfall phase continues more than a few e-folds but less than about 60 e-folds.
That is, the phase transition occurs around the middle between the horizon exit of the scale of the cosmic microwave background (CMB) and the end of inflation.
In such cases, $\phi$'s vev is almost equal to $\phi_c$ during about 60 e-folds and therefore we can Taylor expand the inflaton's potential $V(\phi)$ around $\phi_c$ regardless of the motivating UV theories.
Namely, adopting the notation of ref.~\cite{Clesse:2015wea}, we analyze the following form of the potential.
\begin{align}\label{potential}
	V(\phi,\psi)=\Lambda^4\left[\left(1-\frac{\psi^2}{M^2}\right)^2+2\frac{\phi^2\psi^2}{\phi_c^2M^2}+\frac{\phi-\phi_c}{\mu_1}
	-\frac{(\phi-\phi_c)^2}{\mu_2^2}\right]. 
\end{align}
It has five dimensionful parameters as $\Lambda,\,M,\,\phi_c,\,\mu_1,$ and $\mu_2$. Among them, two d.o.f. can be fixed by the information of 
the amplitude and tilt of the power spectrum of the curvature perturbations on the CMB scale.
In the mild-waterfall case, the CMB scale corresponds with the point in the valley phase, where the waterfall field is still irrelevant due to its large mass.
Therefore the perturbations can be analyzed linearly as the simple single-field slow-roll case.
At first the slow-roll parameters are given by,
\begin{align}
	\epsilon_V=\left.\frac{M_p^2}{2}\left(\frac{V_\phi}{V}\right)\right|_{\phi\sim\phi_c,\psi\sim0}\simeq\frac{M_p^2}{2\mu_1^2}, \quad
	\eta_V=\left.M_p^2\frac{V_{\phi\phi}}{V}\right|_{\phi\sim\phi_c,\psi\sim0}\simeq-\frac{2M_p^2}{\mu_2^2},
\end{align}
where $M_p$ is the reduced Planck mass $\sqrt{8\pi G}^{-1}\simeq2.4\times10^{18}\,\mathrm{GeV}\simeq4.3\times10^{-6}\,\mathrm{g}$.
The spectral index $n_s$ is, in the slow-roll limit,
\begin{align}
	n_s=1-6\epsilon_V+2\eta_V\simeq1-\frac{4M_p^2}{\mu_2^2},
\end{align}
where we assumed that $\eta_V$ dominates $\epsilon_V$ (as can be checked easily for specific parameter regions shown in the following sections),
which is the case for small field inflation.
From this relation, with the Planck's best fit value $n_s\simeq0.9655$~\cite{Ade:2015xua}, $\mu_2$ should be fixed to,
\begin{align}
	\frac{\mu_2}{M_p}=\frac{2}{\sqrt{1-n_s}}\simeq11.
\end{align}
Also the amplitude of the power spectrum is given by,
\begin{align}
	A_s=\frac{1}{24\pi^2M_p^4}\frac{V}{\epsilon_V}\simeq\frac{\Lambda^4\mu_1^2}{12\pi^2M_p^6}.
\end{align}
Again it should be fixed by the Planck's result $A_s\simeq2.198\times10^{-9}$, which gives the following relation:
\begin{align}\label{Lambda by mu1}
	\left(\frac{\Lambda}{M_p}\right)^4\simeq2.198\times10^{-9}\times12\pi^2\left(\frac{\mu_1}{M_p}\right)^{-2}.
\end{align}
In the following sections, we will fix $\Lambda$ with this constraint and take $M,\phi_c$ and $\mu_1$ as free parameters.
 
In ref.~\cite{Clesse:2015wea}, Clesse and Garcia-Bellido (CG) analytically approximated the curvature perturbations during the waterfall phase and estimated the PBH abundance.
At first they calculated the variance of the waterfall field at the critical point, namely $\sigma_\psi^2=\braket{\psi^2}|_{\phi=\phi_c}$, in the stochastic formalism which we will describe in the next section, as,
\begin{align}\label{sigmapsi}
	\sigma_\psi=\left(\frac{\sqrt{2}\Lambda^4 M\phi_c^{1/2}\mu_1^{1/2}}{96\pi^{3/2}M_p^4}\right)^{1/2}.
\end{align}
Then it was used as an initial condition of $\psi$ at the beginning of the waterfall phase, that is $\psi_0=\psi|_{\phi=\phi_c}\simeq\sigma_\psi$.
The curvature perturbations were finally calculated by the standard linear perturbation theory.
We briefly review the results below, while the details are omitted.

With the potential~(\ref{potential}), the slow-roll e.o.m. is given by,
\begin{numcases}
	{}	
	\label{phi's slow-roll eom}
	3H\dot{\phi}=-V_\phi\simeq-\frac{\Lambda^4}{\mu_1}-\frac{4\Lambda^4\psi^2}{M^2\phi_c^2}\phi, \\
	\label{psi's slow-roll eom}
	3H\dot{\psi}=-V_\psi\simeq-\frac{4\Lambda^4}{M^2}\left(\frac{\phi^2}{\phi_c^2}-1\right)\psi,
\end{numcases}
where $V_\phi$ and $V_\psi$ denotes the derivatives of the potential with respect to $\phi$ and $\psi$.
Here we omitted the higher-order terms. Then CG divided the waterfall phase into two stage; in the first phase-1 the second term of the right side of eq.~(\ref{phi's slow-roll eom}) is negligible
and in the phase-2 that dominates over the first term. With the approximation $H^2\simeq\Lambda^4/3M_p^2$, the e-folds for the phase-1 and 2 can be calculated as,
\begin{align}
	N_1\simeq\frac{\sqrt{\chi_2}M\phi_c^{1/2}\mu_1^{1/2}}{2M_p^2}, \quad
	N_2\simeq\frac{M\phi_c^{1/2}\mu_1^{1/2}}{4M_p^2\sqrt{\chi_2}},
\end{align}
where,
\begin{align}\label{chi2}
	\chi_2=\log\left(\frac{\phi_c^{1/2}M}{2\mu_1^{1/2}\psi_0}\right),
\end{align}
gives the $\psi$'s field value at the transition point between phase-1 and 2 by $\psi=\psi_0\ee^{\chi_2}$.
Therefore the total e-folds for the waterfall phase is given by,
\begin{align}\label{Nwater}
	N_\mathrm{water}\simeq N_1+N_2\simeq\left(\frac{\sqrt{\chi_2}}{2}+\frac{1}{4\sqrt{\chi_2}}\right)\frac{M\phi_c^{1/2}\mu_1^{1/2}}{M_p^2}.
\end{align} 

Also, according to the $\delta N$ formalism~\cite{Starobinsky:1986fxa,Salopek:1990jq,Sasaki:1995aw,Sasaki:1998ug,Lyth:2004gb}, 
the power spectrum of the curvature perturbations can, in the linear perturbation theory, be approximated by,
\begin{align}
	\mathcal{P}_\zeta(k)=\frac{k^3}{2\pi^2}\int\dd^3x\braket{\zeta(0)\zeta(\mathbf{x})}\ee^{-i\mathbf{k}\cdot\mathbf{x}}\simeq\left.\frac{H^2}{(2\pi)^2}(N_\phi^2+N_\psi^2)\right|_{aH=k},
\end{align}
where $N_\phi$ and $N_\psi$ are the derivatives of the backward e-folds with respect to $\phi$ and $\psi$ respectively.
Assuming the dominant contribution comes from the variation of the phase-1 e-folds $N_1$ due to the $\psi$'s fluctuations, the power spectrum is given by,
\begin{align}\label{calPzeta}
	\mathcal{P}_\zeta(k)\simeq\frac{\Lambda^4 M^2\phi_c\mu_1}{192\pi^2M_p^6\chi_2\psi_k^2},
\end{align}
with $\psi_k=\psi_0\ee^{\chi_k}$, $\chi_k=4\phi_c\mu_1\xi_k^2/M^2$, and $\xi_k=-M_p^2(N_1+N_2-N_k)/(\phi_c\mu_1)$. 
$N_k$ is the backward e-folds corresponding with
considered comoving scale $k$, namely $k=\ee^{-N_k}k_f$ where $k_f$ is the comoving horizon scale $aH$ at the end of inflation.
The power spectrum is maximal at the critical point as,
\begin{align}\label{Pzmax}
	\mathcal{P}_{\zeta,\mathrm{max}}\simeq\frac{\Lambda^4 M^2\phi_c\mu_1}{192\pi^2M_p^6\chi_2\psi_0^2}=\frac{M\phi_c^{1/2}\mu_1^{1/2}}{2\sqrt{2\pi}M_p^2\chi_2}.
\end{align}

There are two key points in these results. The first one is that the power spectrum given by CG has its maximum exactly at the critical point. 
However, in our calculation, the power spectrum is maximized slightly after the critical point 
as shown in section~\ref{Example power spectrum} because the quantum fluctuations of the waterfall field themselves become larger after the critical point due to its tachyonic mass.

As the second and much more important point, both of the e-folding numbers for the waterfall phase $N_\mathrm{water}$ and the maximum of the power spectrum $\mathcal{P}_{\zeta,\mathrm{max}}$
depend almost only on the specific parameter combination $M^2\phi_c\mu_1/M_p^4$ called $\Pi^2$ by CG, except for the small logarithmic dependence due to $\chi_2$.
Indeed, from eqs.~(\ref{Nwater}) and (\ref{Pzmax}), we can easily find a one-to-one monotonic increase correspondence between $N_\mathrm{water}$ and $\mathcal{P}_\zeta$.
Before that, let us clarify the typical value of $\chi_2$. Substituting the initial condition $\psi_0=\sigma_\psi$ (\ref{sigmapsi}) into eq.~(\ref{chi2}) and using the CMB normalization~(\ref{Lambda by mu1}),\footnote{
Even if we do not use the CMB normalization and deal with $\Lambda$ as a free parameter, $\chi_2$ depends on $\Lambda$ only logarithmically.}
$\chi_2$ can be simply written as,
\begin{align}
	\chi_2=\log\left(\left(\frac{2}{\pi}\right)^{1/4}A_s^{-1/2}\,\Pi^{1/2}\right), \quad \Pi^2=\frac{M^2\phi_c\mu_1}{M_p^4},
\end{align}
where $A_s=2.198\times10^{-9}$. From this expression, it can been seen that $\chi_2$ is around 10 for typical values $10\ltsim\Pi^2\ltsim1000$ in the mild-waterfall cases. 
Therefore, from eqs.~(\ref{Nwater}) and (\ref{Pzmax}),
we can obtain the following relation, which does not depend on any detail parameterization.
\begin{align}
	\mathcal{P}_{\zeta,\mathrm{max}}\simeq\frac{1}{\sqrt{2\pi\chi_2^3}}N_\mathrm{water}\simeq0.01N_\mathrm{water}.
\end{align}
Here we neglected the $\chi_2^{-1/2}$ term in eq.~(\ref{Nwater}). Since the PBH constraints on the curvature perturbations are $\mathcal{P}_\zeta\ltsim\mathcal{O}(0.01)$ as we will mention in the next section,
it can be obviously seen that the PBH overproduction is inevitable in the mild-waterfall cases such that $N_\mathrm{water}\gtsim\mathcal{O}(10)$. In the next section, we will check and clarify this estimation
with use of the stochastic formalism.

\section{Parameter search}\label{Parameter search}
Though we reviewed the result in the linear perturbation theory in the previous section, the dynamics of the waterfall field around the critical point is actually dominated by the Hubble fluctuations.
Therefore the linear perturbation with respect to $\psi$ around the critical point essentially breaks down (c.f. \cite{Martin:2011ib,Levasseur:2013tja}).
Accordingly we calculate the curvature perturbations without the perturbative expansions with respect to $\phi$ and $\psi$ with use of the stochastic formalism.
In this section we introduce the stochastic formalism at first. Then we calculate the curvature perturbations in the wide parameter region.
From their results, we conclude that PBHs are overproduced in the mild-waterfall cases as a main claim of this paper.

\subsection{Stochastic formalism}
The stochastic formalism was proposed by Starobinsky in 1986~\cite{Starobinsky:1986fx} (see also \cite{Nambu:1987ef,Kandrup:1988sc,Nakao:1988yi,Mollerach:1990zf,Linde:1993xx,Starobinsky:1994bd}). 
In this formalism, the superhorizon coarse-grained fields, namely,
\begin{align}\label{phiIR}
	\phi_\mathrm{IR}(t,\mathbf{x})=\int\frac{\dd^3k}{(2\pi)^3}W\left(\frac{k}{\epsilon aH}\right)\ee^{i\mathbf{k}\cdot\mathbf{x}}\phi_\mathbf{k}(t),
\end{align}
are treated as the classical background fields. Here $W(k/k_s)$ is a window function and $\epsilon$ is a small positive parameter. As a window function, the simple step function $\theta(\epsilon aH-k)$ is often used 
for brevity. $\epsilon$ divides the scalar fields into the classicalized part and the quantum part. That is, the modes for $k<\epsilon aH$ are well classicalized and can be treated as classical fields, while
the modes for $k>\epsilon aH$ should be assumed to be the quantum operators. 
To take sufficiently superhorizon modes and also validate the perturbative expansions with respect to $\epsilon$ which we use below,
$\epsilon$ should be less than unity. In this paper the value of 0.01 is mainly used for $\epsilon$, while several results for $\epsilon=0.1$ are also shown to see the $\epsilon$-dependences.

For the above coarse-grained fields, the e.o.m. reads~\cite{Morikawa:1989xz},
\begin{align}\label{IR eom}
	\begin{cases}
		\displaystyle
		\dif{\phi_{\mathrm{IR},i}}{N}=\frac{\pi_{\mathrm{IR},i}}{H}+\mathcal{P}_{\phi_i}^{1/2}(k=\epsilon aH)\xi_i, \\[5pt]
		\displaystyle
		\dif{\pi_{\mathrm{IR},i}}{N}=-3\pi_{\mathrm{IR},i}-\frac{V_i}{H},
	\end{cases}
\end{align}
in the leading order with respect to $\epsilon$. $\mathcal{P}_{\phi_i}^{1/2}$ denotes the power spectrum of $\phi_i$, i.e. $\mathcal{P}_{\phi_i}(k)=\frac{k^3}{2\pi^2}\int\dd^3x\braket{\phi_i(0)\phi_i(\mathbf{x})}
\ee^{-i\mathbf{k}\cdot\mathbf{x}}$. The subscript $i$ labels the flavors of the scalar fields for the multi-field cases. Without the term of $\xi_i$, it is recovered to the standard e.o.m. for the homogenous fields.
This $\xi$ term as an important difference is interpreted as a classical Gaussian random variable having following stochastic properties.\footnote{Here we assume there is no correlation between the different flavors of $\xi$.
However the correlations between them due to the interactions before the horizon exit can be also included. In this paper, we omit them for simplicity after easily checked that they do not affect the result at all.}
\begin{align}
	\begin{cases}
		\braket{\xi_i(N,\mathbf{x})}=0,  \\
		\displaystyle
		\braket{\xi_i(N,\mathbf{x})\xi_j(N^\prime,\mathbf{x}^\prime)}=\delta_{ij}\frac{\sin(\epsilon aHr)}{\epsilon aHr}\delta(N-N^\prime), \quad r=|\mathbf{x}-\mathbf{x}^\prime|.
	\end{cases}
\end{align}
For the coarse-grained fields, $\frac{\sin(\epsilon aHr)}{\epsilon aHr}$ can be approximated by $\theta(1-\epsilon aHr)$. The delta function type property for the time variable comes from the fact that
we used the sharp window function for eq.~(\ref{phiIR}). Anyway it represents the fact that \emph{the superhorizon coarse-grained fields
receive the Gaussian white noise independent for each Hubble patch and its amplitude is given by the scalar fields' perturbations 
$\mathcal{P}_{\phi_i}^{1/2}$}. This noise term comes from the inflow of the UV part $\phi_\mathrm{UV}=\phi-\phi_\mathrm{IR}$ into the IR part
for every time. As we indicated, the UV part originally behaves as a quantum field but it is redshifted and classicalized at the time of $k=\epsilon aH$
to join in the IR part. At this time the exact field value of this joining mode cannot be determined due to its quantum property. Instead it is 
interpreted as a classical random variable and its amplitude can be calculated in quantum field theory as $\mathcal{P}_{\phi_i}$ of course.

Due to the noise term, every Hubble patch is assumed to evolve independently in the stochastic formalism.
In this sense, the all background parameter values in eq.~(\ref{IR eom}), namely $H,\,V_i$, and $\mathcal{P}_{\phi_i}$, should be 
determined in each Hubble patch by the scalar field values of that patch. Inversely, if one concentrate on the dynamics of one Hubble patch,
the e.o.m. reduces to the following self-closed Langevin equations.
\begin{align}\label{langevin}
	\begin{cases}
		\displaystyle
		\dif{\phi_i}{N}(N)=\frac{\pi_i}{H}(N)+\mathcal{P}_{\phi_i}^{1/2}(N)\xi_i(N), \\[5pt]
		\displaystyle
		\dif{\pi_i}{N}(N)=-3\pi_i(N)-\frac{V_i}{H}(N), \\[5pt]
		V_i(N)=V_i(\phi_1(N),\phi_2(N),\cdots),  \\[5pt]
		\displaystyle
		3M_p^2H^2(N)=\sum_i\frac{\pi_i^2}{2}+V(\phi_1(N),\phi_2(N),\cdots),  \\
		\braket{\xi_i(N)}=0, \\
		\braket{\xi_i(N)\xi_j(N^\prime)}=\delta_{ij}\delta(N-N^\prime).
	\end{cases}
\end{align}
However in regards to $\mathcal{P}_{\phi_i}$, one should calculate the dynamics of all subhorizon modes with the above Langevin eq. 
to obtain the value of them, strictly speaking. In this paper we approximate them by the constant mass solution as,
\begin{align}\label{calPphi}
	\mathcal{P}_{\phi_i}(k=\epsilon aH)=\frac{H^2}{8\pi}\epsilon^3|H_{\nu_i}^{(1)}(\epsilon)|^2,
\end{align}
where $H_\nu^{(1)}(x)$ is the Hankel function of the first kind given by,
\begin{align}
	H_\nu^{(1)}(x)=J_\nu(x)+iY_\nu(x),
\end{align}
with the Bessel functions of the first and second kind, $J_\nu(x)$ and $Y_\nu(x)$, and $\nu_i$ is defined by,
\begin{align}
	\nu_i=\sqrt{\frac{9}{4}-\frac{V_{ii}}{H^2}},
\end{align}
for $\frac{V_{ii}}{H^2}\leq\frac{9}{4}$. For massive fields as $\frac{V_{ii}}{H^2}>\frac{9}{4}$, we simply assume that their Hubble noise vanishes.
Since numerical calculations of the Bessel functions are time consuming, we use the asymptotic forms of them for small arguments as,
\begin{align}
	J_\nu(x)\simeq\frac{1}{\Gamma(\nu+1)}\left(\frac{x}{2}\right)^\nu, \quad Y_\nu(x)\simeq-\frac{\Gamma(\nu)}{\pi}\left(\frac{2}{x}\right)^\nu.
\end{align}
In the following sections, we numerically solve these equations simultaneously.

\subsection{Mean and variance of e-folds}
As we saw in the previous section, the stochastic formalism gives the e.o.m. for the scalar fields coarse-grained on the horizon scale.
Therefore the dynamics of one spatial point in the stochastic formalism can be regarded as that of one Hubble patch.
Namely, in the stochastic formalism, the scalar field in each Hubble patch behaves as a Brownian motion drifted by the potential force.
On the other hand, according to the $\delta N$ formalism~\cite{Starobinsky:1986fxa,Salopek:1990jq,Sasaki:1995aw,Sasaki:1998ug,Lyth:2004gb}, 
the gauge invariant curvature perturbations on the superhorizon scale are given by the difference of the e-folds between 
the initial flat slice and the final uniform density slice. That is, since the dynamics of each Hubble patch automatically fluctuates due to the noise, the e-folding numbers also vary over the universe and 
their fluctuations are nothing but the curvature perturbations. Strictly speaking, the obtained curvature perturbations are coarse-grained values on the horizon scale at the end of inflation.

Let us describe the method more concretely. At first, one must determine the initial flat slice in the valley phase and the final uniform density slice around the end of inflation.
Here, regarding the initial flat slice, note that in the valley phase inflation can be approximated as the single-field case and moreover the curvature perturbations are much smaller than those expected in the waterfall phase.
Therefore, neglecting the curvature perturbations, the initial flat slice can be approximated by the uniform $\phi$ slice and the $\psi$'s field value is almost irrelevant. 
Next, making many realizations of the Langevin equations from the initial field values to the final energy density value, one can obtain various realizations of the e-folding numbers. 
Their deviations from the mean value $\braket{N}$ are nothing but the data set of the coarse-grained curvature perturbations. Though the information of the correlation function like $\braket{\zeta(\mathbf{x}_1)\zeta(\mathbf{x}_2)}$
for $\mathbf{x}_1\neq\mathbf{x}_2$ cannot be derived at this time, at least the probability distribution function (PDF) of the coarse-grained curvature perturbations can be obtained up to the realization errors.
With use of this PDF, one can calculate the formation rate of PBHs whose masses are larger than that corresponding with the horizon scale at the end of inflation.
In this section, let us roughly estimate the PBH abundance by this quantity and find the parameter constraints.

\begin{figure}
	\centering
	\begin{tabular}{cc}
		\begin{minipage}{0.490541\hsize}
			\centering
			\includegraphics[width=\hsize]{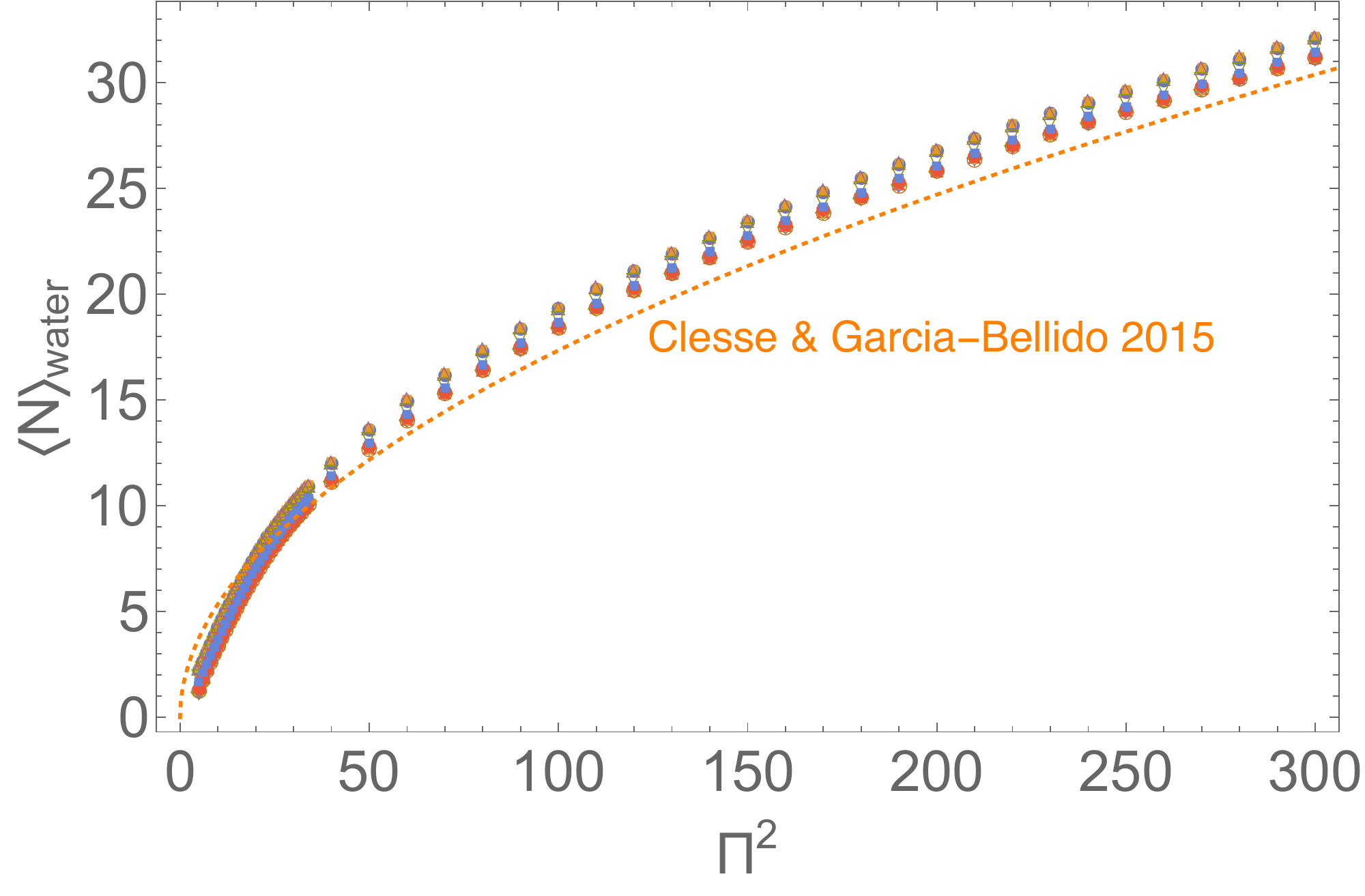}
		\end{minipage}
		\begin{minipage}{0.509459\hsize}
			\centering
			\includegraphics[width=\hsize]{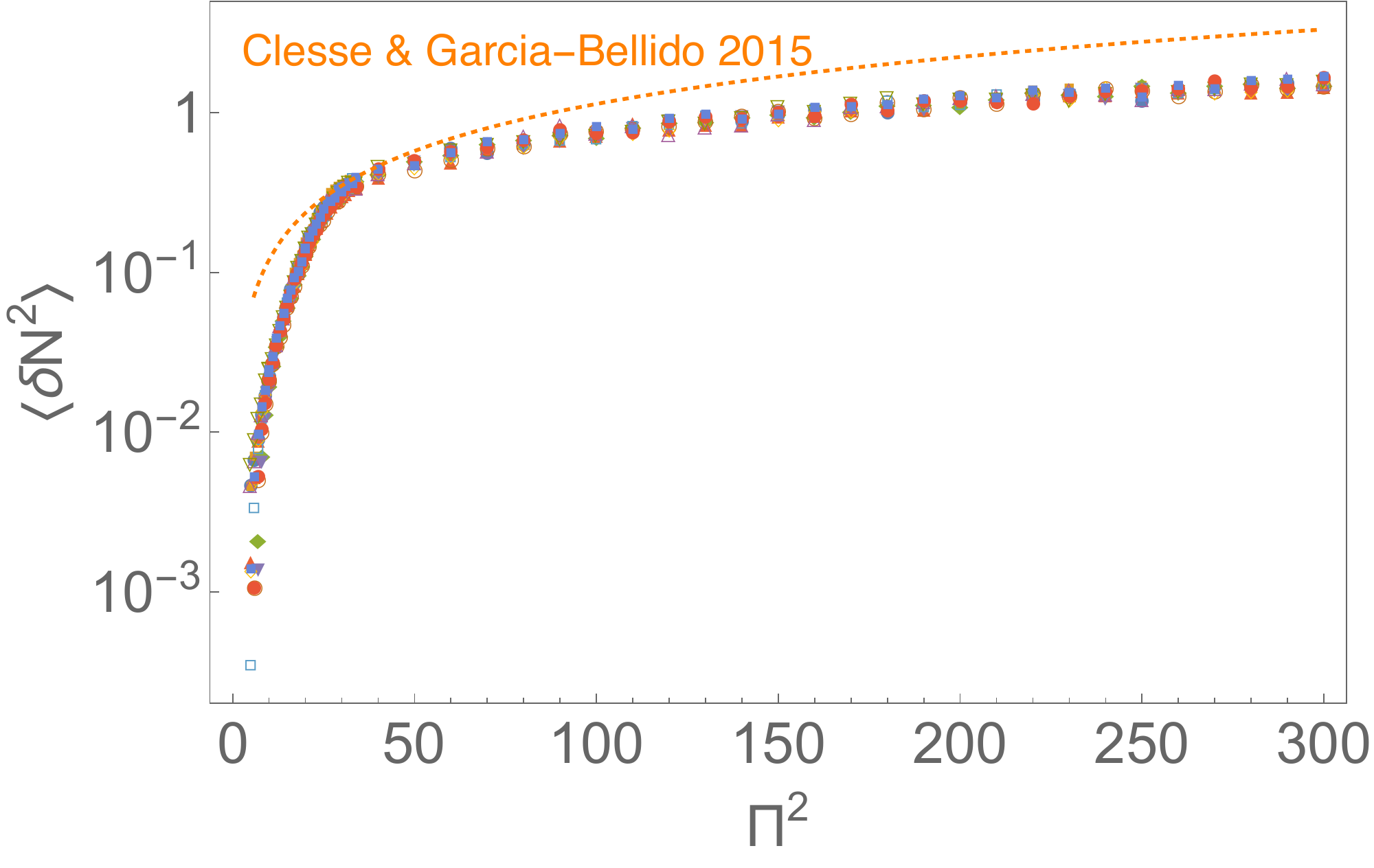}
		\end{minipage}
	\end{tabular}
	\caption{
	The mean e-folds of the waterfall phase (left panel) and the variance of their perturbations (right panel) vs. 
	$\Pi^2=M^2\phi_c\mu_1/M_p^4$ for various parameter sets in the searching region~(\ref{searching region}). 
	$\mu_1$ is varied for each set of $M$ and $\phi_c$. There are 12 sets of $(M,\,\phi_c)$ represented by different markers although they cannot distinguished in the figure. 
	2000 realizations are made for 
	each data point. It is clearly shown that both of $\braket{N}$ and $\braket{\delta N^2}$ depend almost only on $\Pi^2$ as 
	ref.~\cite{Clesse:2015wea} suggested. However, while their results which are represented by orange dotted lines are well consistent
	with our calculations for $\braket{N}$, there are factor differences in $\braket{\delta N^2}$. Anyway these plots indicate $\Pi^2$ should be less than about 10 to satisfy the PBH constraint $\braket{\delta N^2}\ltsim0.01$
	and it means the waterfall phase cannot continue more than about 5 e-folds.
	}
	\label{N and dN2}
\end{figure}

For parameter search, we used the following three searching regions.
\begin{align}\label{searching region}
	\begin{array}{lll}
		\bullet & 10^{-4}M_p\le M\le10^{-1}M_p, \quad & \phi_c=\sqrt{2}M, \quad \text{(SUSY like assumption)} \\
		\bullet & M=0.1M_p, \quad & 10^{-4}M_p\le \phi_c\le10^{-1}M_p, \\
		\bullet & 10^{-4}M_p\le M\le10^{-1}M_p, \quad & \phi_c=0.1M_p.  
	\end{array}
\end{align}
$\mu_1$ is also varied so that $\Pi^2=M^2\phi_c\mu_1/M_p^4$ takes the value up to 300. 
$\Lambda$ is given by eq.~(\ref{Lambda by mu1}) for each value of $\mu_1$.
In figure~\ref{N and dN2}, we plot the mean e-folds for the waterfall phase $\braket{N}_\mathrm{water}$ and the variance of their perturbations $\braket{\delta N^2}=\braket{N^2}-\braket{N}^2$ 
vs. $\Pi^2=M^2\phi_c\mu_1/M_p^4$ for various parameters in the above searching region.
$\mu_1$ is varied for each parameter set $(M,\,\phi_c)$. 
Also the CG's analytic results~(\ref{Nwater}) and (\ref{calPzeta}) are shown as orange dotted lines. Here note that the variance and the power spectrum are related by,
\begin{align}
	\braket{\delta N^2}\simeq\int^{\braket{N}_\mathrm{water}}_0\mathcal{P}_\zeta(k)\dd N_k.
\end{align}
From this figure, it is found that the mean e-folds $\braket{N}$ obtained in the stochastic formalism is well fitted by the CG's result.
On the other hand, there are factor differences in the variance $\braket{\delta N^2}$.
It clearly shows the non-perturbative effects which CG did not include.
However at least it can be said that the full result of $\braket{\delta N^2}$ also depends only on the specific parameter combination $\Pi^2=M^2\phi_c\mu_1/M_p^4$.
Having in mind that the variance of the curvature perturbations should be roughly less than $10^{-2}$ at most not to overproduce PBHs as we will show in the next subsection,
it can be seen that $\Pi^2$ should be less than around 10, which indicates the waterfall phase cannot continue more than about 5 e-folds.

\begin{figure}
	\centering
	\begin{tabular}{cc}
		\begin{minipage}{0.489879\hsize}
			\centering
			\includegraphics[width=\hsize]{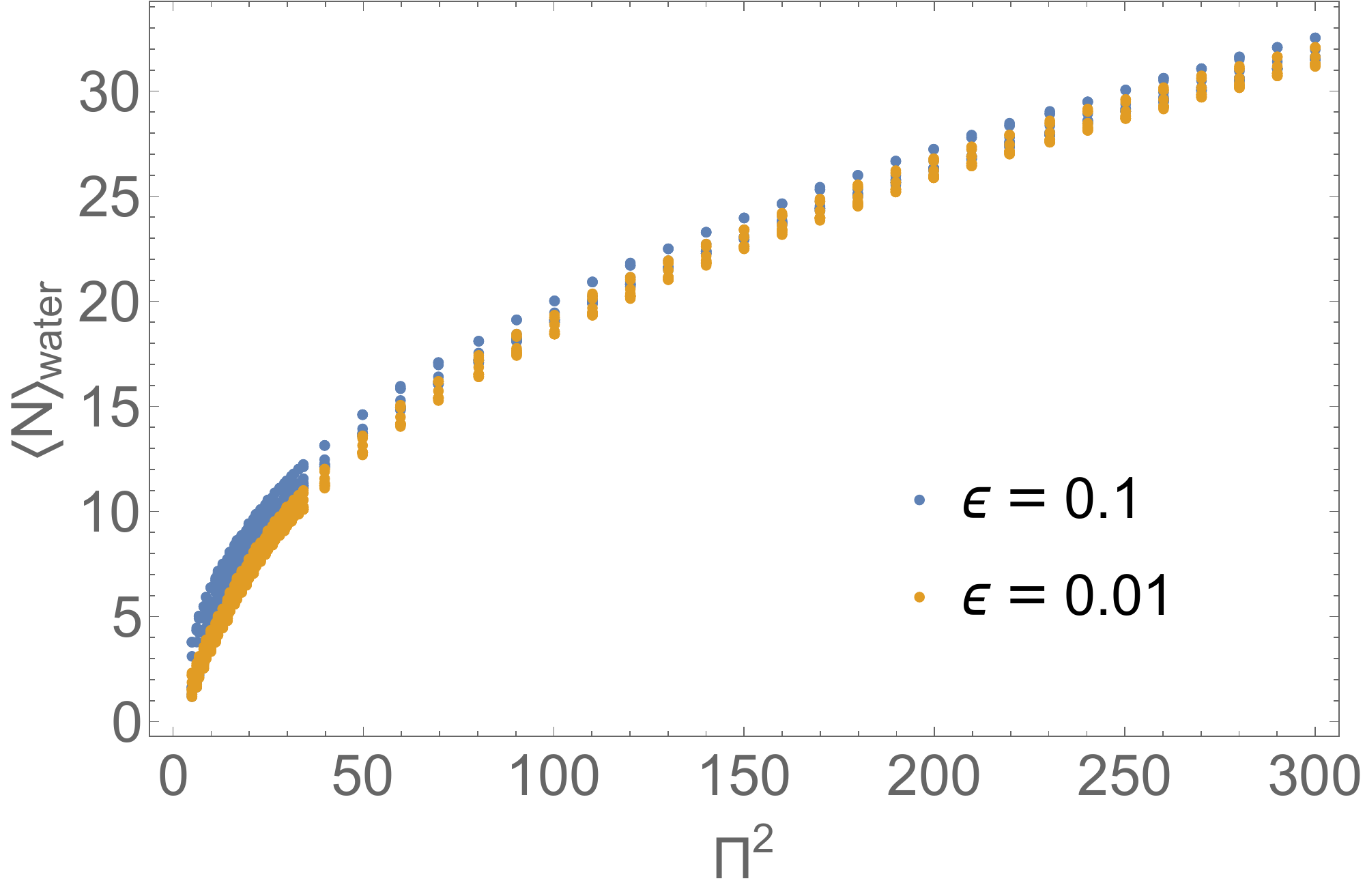}
		\end{minipage}
		\begin{minipage}{0.510121\hsize}
			\centering
			\includegraphics[width=\hsize]{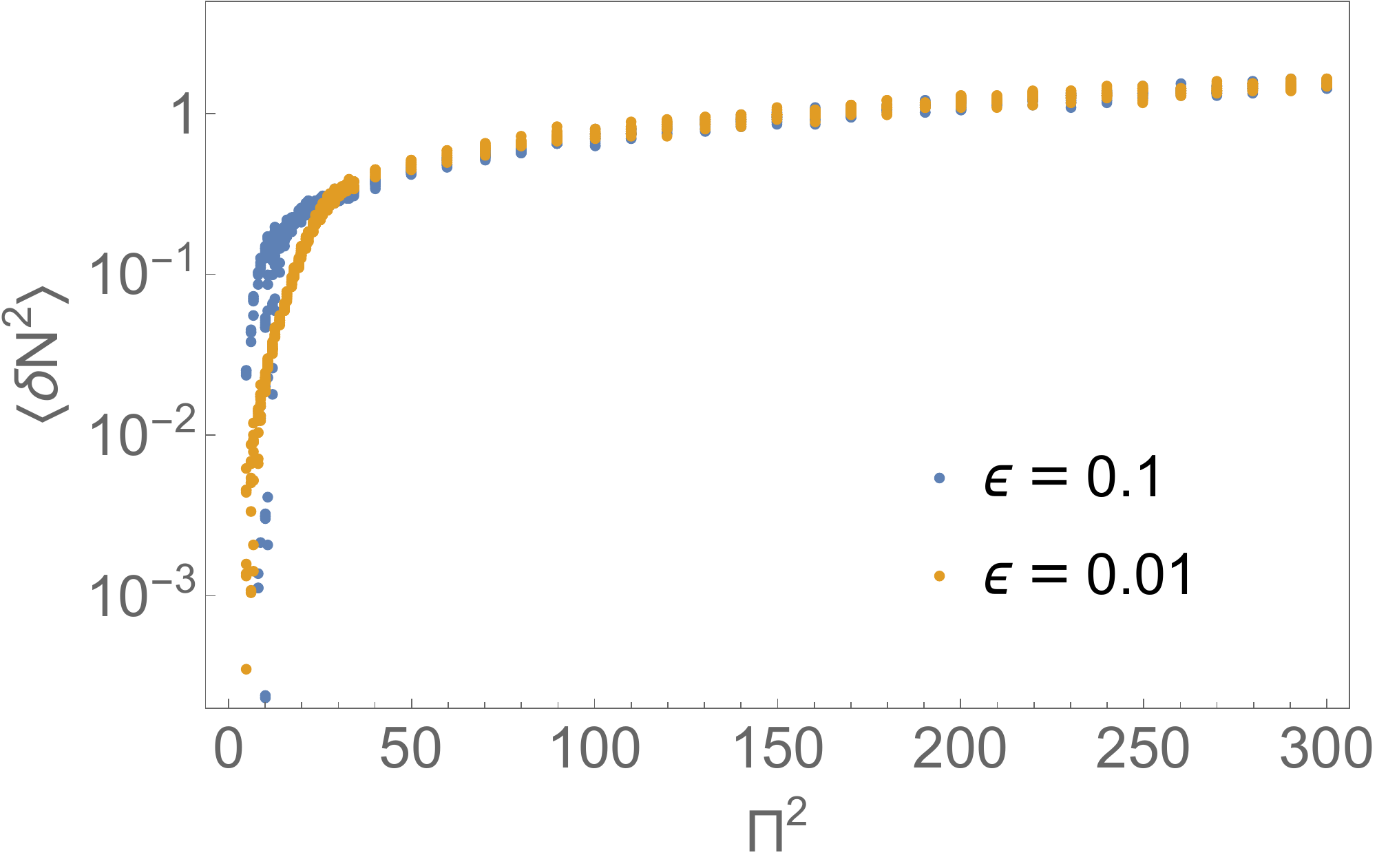}
		\end{minipage}
	\end{tabular}
	\caption{The same plots to figure~\ref{N and dN2} for $\epsilon=0.1$ and $0.01$. The orange datas are the same ones of figure~\ref{N and dN2}.
	While $\braket{N}_\mathrm{water}$ has good agreement between different $\epsilon$, $\braket{\delta N^2}$ shows a large $\epsilon$-dependence in low $\Pi^2$ which
	indicates that the results in the stochastic formalism for low $\Pi^2$ might be unreliable.}
	\label{N and dN2 epsilon}
\end{figure}

Here let us mention the $\epsilon$-dependence of the results. As we said, the stochastic formalism has an indeterminate parameter $\epsilon$ which fixes the separation between the classical superhorizon and
the quantum subhorizon modes. Since this is just the uncertainty of the formalism, any result should have little $\epsilon$-dependence for reliable calculations. 
To see the $\epsilon$-dependence, we also show $\braket{N}$ and $\braket{\delta N^2}$ for $\epsilon=0.1$
in figure~\ref{N and dN2 epsilon}, comparing them to those for $\epsilon=0.01$. It shows that $\braket{\delta N^2}$ has relatively large differences in low $\Pi^2$.
For low $\Pi^2$, the waterfall phase does not continue so long as already shown. On the other hand, for small $\epsilon$, the modes shorter than the coarse-graining scale $k=\epsilon aH$ are 
erased in the stochastic formalism. Specifically, since $-\log0.01\simeq4.6$, the perturbations generated in about last 4.6 e-folds cannot be treated in the calculations where $\epsilon=0.01$. 
Therefore, for low $\Pi^2$, the contribution of the perturbations after the critical point cannot be taken into account well in the case of small $\epsilon$, and that is the reason why $\braket{\delta N^2}$
for $\epsilon=0.01$ is suppressed compared to that for $\epsilon=0.1$.
Anyway the case of low $\Pi^2$ is slightly out of the range of application of the stochastic formalism, but it does not change the results for large $\Pi^2$ and the main conclusion that massive PBHs are overproduced.

\subsection{PBH abundance}
In this subsection, we estimate the PBH abundance including the non-Gaussian (NG) effects.
Following the Press-Schechter approach~\cite{Press:1973iz}, if the coarse-grained curvature perturbations $\zeta_s$ follow the PDF $P(\zeta_s)$, the formation rate of PBHs which are more massive than the mass corresponding with
the coarse-graining scale is given by,
\begin{align}
	\beta(>M_\mathrm{PBH})=2\int^\infty_{\zeta_c}P(\zeta_s)\dd\zeta_s,
\end{align}
where $\zeta_c$ is the threshold for the PBH formation. The factor 2 is conventional to include the effects of mergers and accretions.
For the case of PBHs, the mass $M_\mathrm{PBH}$ is related to the coarse-graining comoving scale $R_s$ by,\footnote{
Here we approximate the PBH mass simply by the horizon mass $M_H$, but it is known to scale, depending on 
the value of the density perturbation $\delta$, as,
\begin{align}
	M_\mathrm{PBH}=k M_H(\delta-\delta_c)^\gamma,
\end{align}
where $\delta_c$ represents the critical threshold value for $\delta$, and $k$ and $\gamma$ are 
some numerical factors~\cite{Choptuik:1992jv,Niemeyer:1997mt,Kuhnel:2015vtw}.
However this effect basically shifts the PBH mass to smaller one and therefore it will not change our main result that
massive PBHs cannot be produced with the appropriate abundance in the mild-waterfall hybrid inflation.}
\begin{align}\label{MPBH}
	M_\mathrm{PBH}\simeq\frac{M_p^2}{H_\mathrm{inf}}(k_fR_s)^2\simeq\frac{M_p^2}{H_\mathrm{inf}}\ee^{2N_s}
	=1.0\times10^4\,\mathrm{g}\left(\frac{H_\mathrm{inf}}{10^9\,\mathrm{GeV}}\right)^{-1}\ee^{2N_s}.
\end{align}
Here $H_\mathrm{inf}$ is the Hubble parameter at the end of inflation and $N_s$ denotes 
the corresponding backward e-folds, $R_s^{-1}=k_f\ee^{-N_s}$. 
We assumed that reheating occurs soon after inflation.
If the power spectrum has a large peak which will be validated in the next section, 
PBHs are formed almost only on the peak scale and the above $\beta$ with the coarse-graining scale being the peak scale can be directly used to be constrained.

\begin{figure}
	\centering
	\begin{tabular}{cc}
		\begin{minipage}{0.498603\hsize}
			\centering
			\includegraphics[width=\hsize]{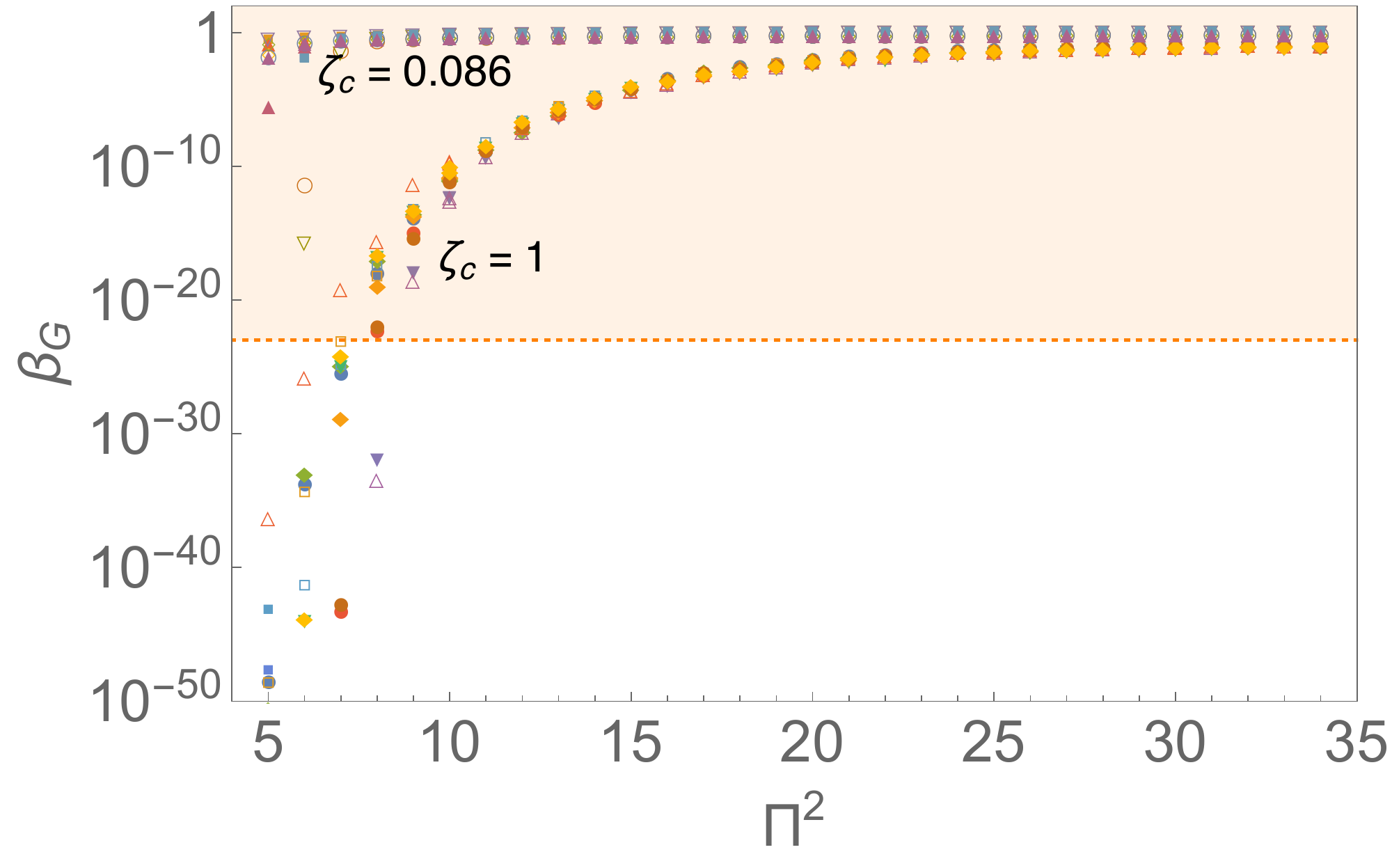}
		\end{minipage}
		\begin{minipage}{0.501397\hsize}
			\centering
			\includegraphics[width=\hsize]{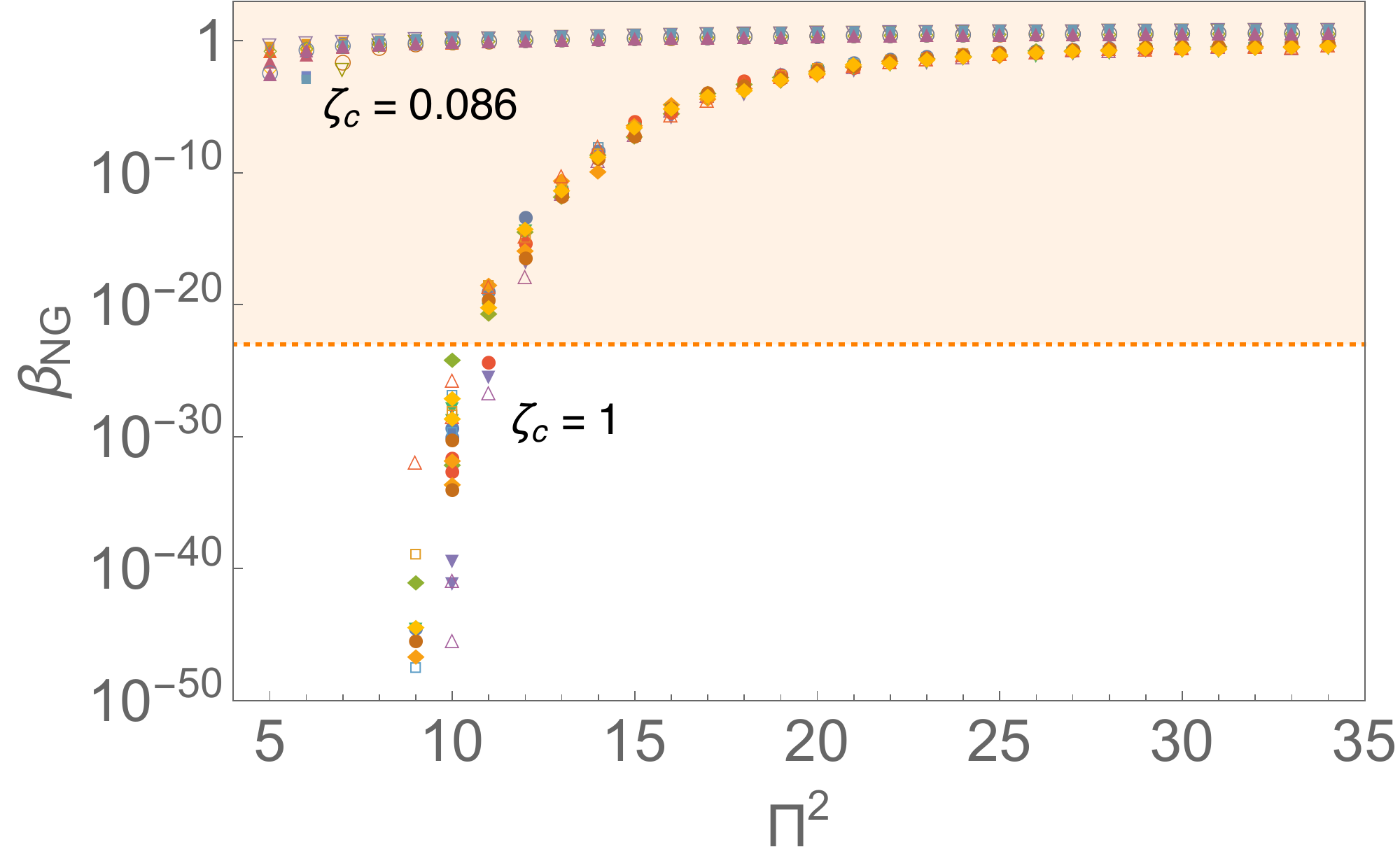}
		\end{minipage}
	\end{tabular}
	\caption{Left panel: the plot of the PBH abundance in the Gaussian assumption $\beta_G$~(\ref{betaG}) 
	with use of the variance shown in figure~\ref{N and dN2}. 
	Namely it includes all contributions of PBHs more massive than the smallest mass scale $M_p^2/H_\mathrm{inf}$ which corresponds with the scale
	at the end of inflation.
	The bottom group is for $\zeta_c=1$ while the upper group is 
	for $\zeta_c\simeq0.086$~\cite{Harada:2013epa}. Also the orange dotted line represents the typical constraints 
	for light PBHs~($\ltsim10^{15}\,\mathrm{g}$), namely $\beta\ltsim10^{-23}$. 
	If $\zeta_c\simeq0.086$, there is no appropriate value of $\Pi^2$ with which PBHs are
	not overproduced. On the other hand, if $\zeta_c=1$, the PBH constraints indicate $\Pi^2\ltsim8$, which means the waterfall phase can
	continue few e-folds as shown in figure~\ref{N and dN2}. 
	Right panel: the same plot with NG corrections~(\ref{betaNG}). Though the results for $\zeta_c\simeq0.086$
	are hardly different, the PBH abundance for $\zeta_c=1$ is suppressed for low $\Pi^2$ compared to the value without NG corrections.
	Then the constraints are slightly weakened to $\Pi^2\ltsim11$ but the duration of the waterfall phase is still as short as 4--5 e-folds. 
	}
	\label{beta and betaNG}
\end{figure}

If one assumes the curvature perturbations follow the Gaussian distribution, $\beta$ can be easily estimated by,\footnote{Though we used 
the variance of the coarse-grained curvature perturbations here, the authors of \cite{Young:2014ana} claimed that the power spectrum
on the considered scale should be used instead of the variance. 
That is because the curvature perturbations are undamped quantities even on superhorizon scales and therefore the variance includes the much
superhorizon modes, which should not affect the PBH formation. 
However now the power spectrum has a large peak as we will show in the next section, 
and the larger scale modes than the peak scale are already suppressed.
Therefore using the variance will not overestimate the PBH abundance so much.
Since we would like to include the NG effects by the form of the third and forth moment, 
namely $\braket{\delta N_s^3}$ and $\braket{\delta N_s^4}$, we used the variance $\braket{\delta N_s^2}$ instead of the power spectrum.} 
\begin{align}\label{betaG}
	\beta_G=2\int_{\zeta_c}\frac{1}{\sqrt{2\pi\sigma_s^2}}\ee^{-\zeta_s^2/2\sigma_s^2}\dd\zeta_s,
\end{align}
where $\sigma_s^2$ denotes the variance of the coarse-grained curvature perturbations, namely, with use of some window function $W(kR_s)$,
\begin{align}
	\sigma_s^2=\braket{\zeta_s^2}=\int W^2(kR_s)\mathcal{P}_\zeta(k)\dd\log k.
\end{align}
On the left panel of figure~\ref{beta and betaNG}, we plot the $\beta_G$ with use of $\braket{\delta N^2}$ shown in figure~\ref{N and dN2} as 
$\sigma_s^2$ for different two threshold values, that is, the simple assumption $\zeta_c=1$
and the recent analytic prediction by Harada et al~\cite{Harada:2013epa}, 
$\zeta_c=\left.\frac{1}{3}\log\frac{3(\chi_a-\sin\chi_a\cos\chi_a)}{2\sin^3\chi_a}\right|_{\chi_a
=\pi\sqrt{\omega}/(1+3\omega)}\simeq0.086$ where $\omega$ is the e.o.s. for radiation, $p/\rho=1/3$.
Also we show the typical constraints for light PBHs $\beta_G\sim10^{-23}$ as an indicator~\cite{Carr:2009jm}.\footnote{Though we want massive PBHs 
$\gtsim10^{15}\,\mathrm{g}$, we will find such massive ones cannot be produced with proper abundance 
in hybrid inflation and then we used the constraints
for light PBHs $\ltsim10^{15}\,\mathrm{g}$.}
Incidentally this value corresponds with 10 sigma rarity, i.e. $\zeta_c\sim10\sigma_s$. Therefore this constraint roughly indicates $\mathcal{P}_\zeta\sim\sigma_s^2\ltsim0.01$ for $\zeta_c=1$ as mentioned previously.
The figure shows $\Pi^2$ should be less than around 8 for the case $\zeta_c=1$, which means the waterfall phase can continue few e-folds.
Also there is almost no proper parameter set with which PBHs are not overproduced if $\zeta_c\simeq0.086$.

\begin{figure}
	\centering
	\begin{tabular}{cc}
		\begin{minipage}{0.502008\hsize}
			\centering
			\includegraphics[width=\hsize]{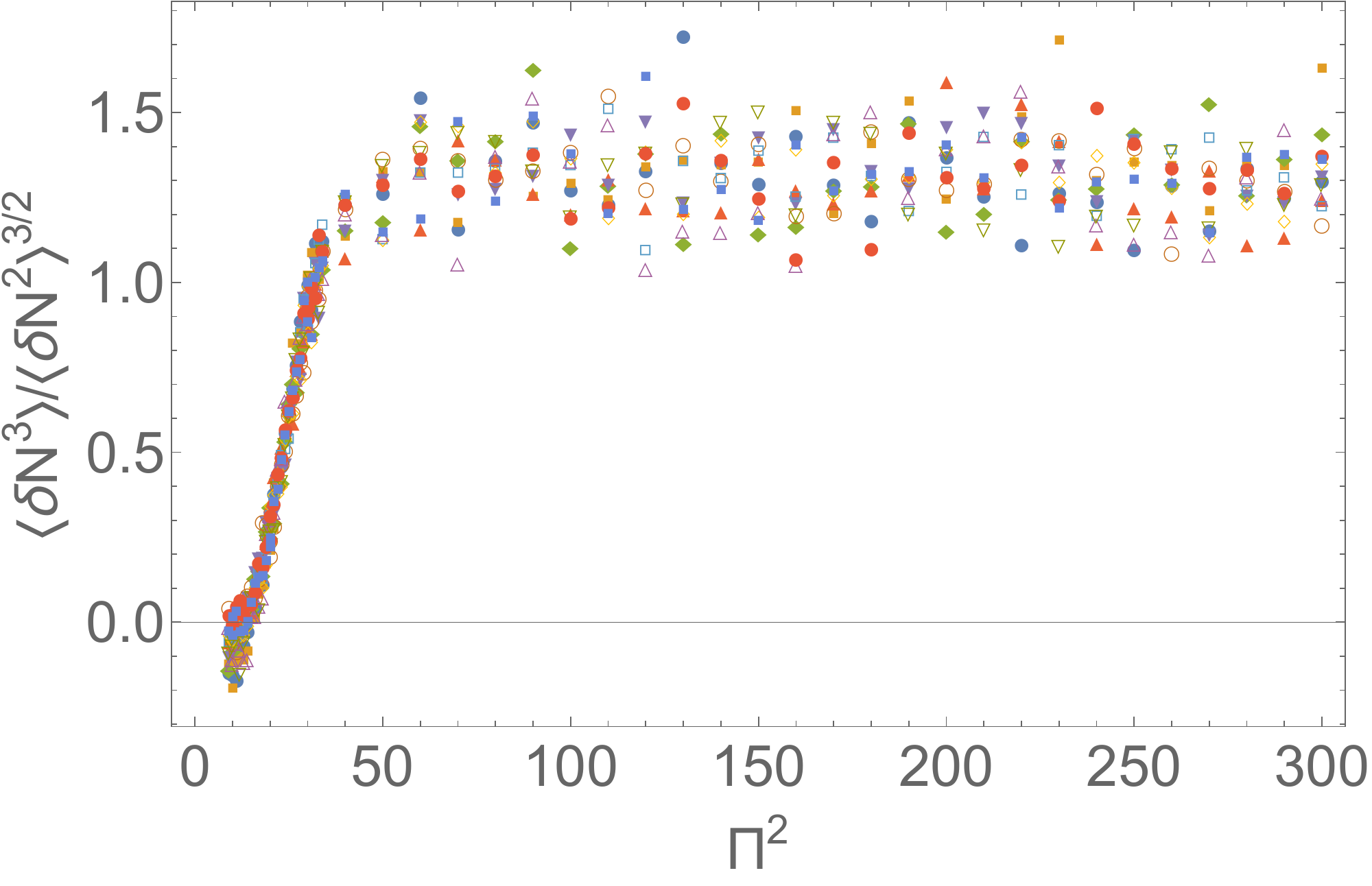}
		\end{minipage}
		\begin{minipage}{0.497992\hsize}
			\centering
			\includegraphics[width=\hsize]{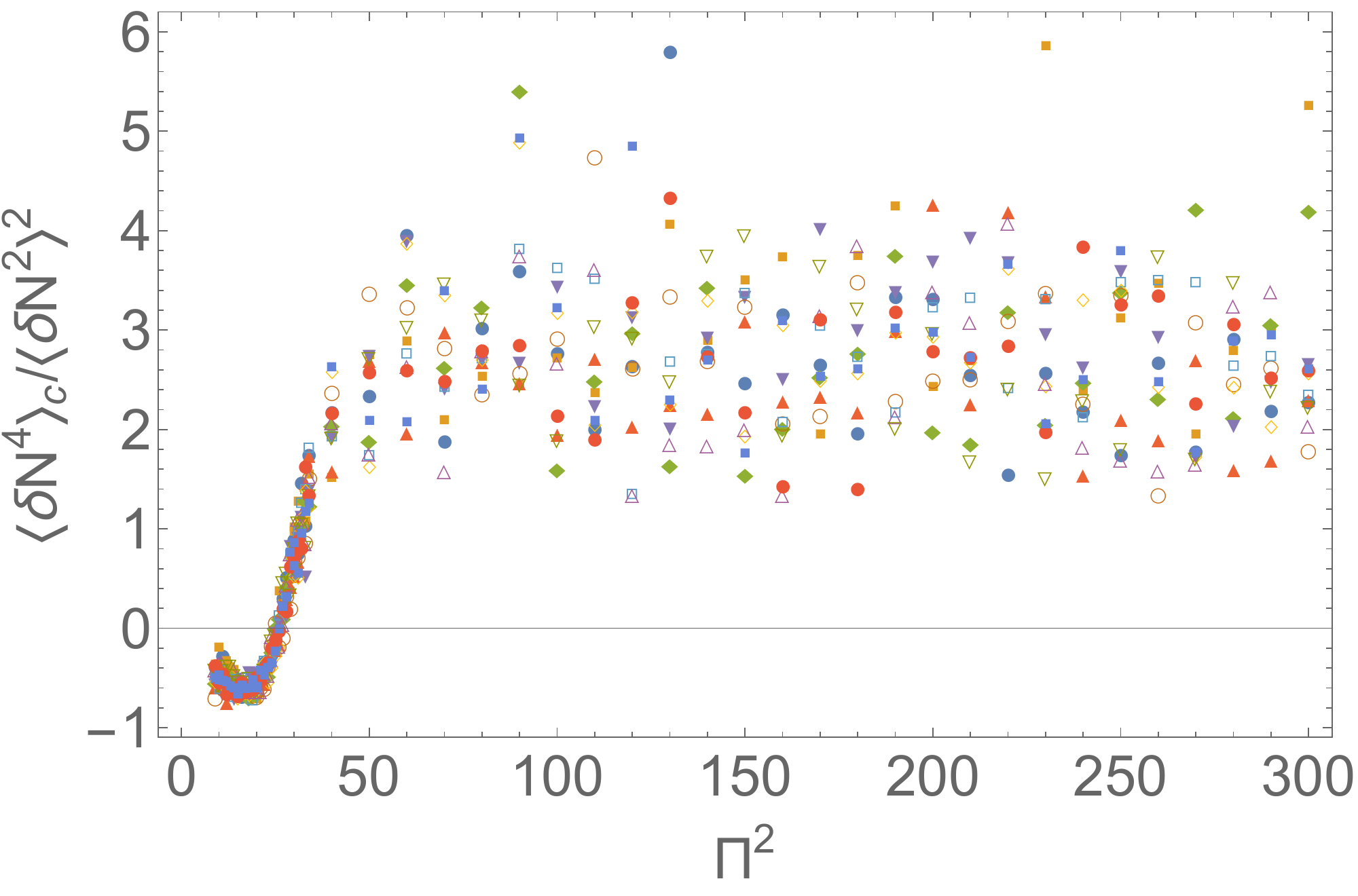}
		\end{minipage}
	\end{tabular}
	\caption{
	The skewness $S^{(3)}=\braket{\delta N^3}/\braket{\delta N^2}^{3/2}$ 
	and kurtosis $S^{(4)}=\braket{\delta N^4}_c/\braket{\delta N^2}^2$ where $\braket{\delta N^4}_c=\braket{\delta N^4}-3\braket{\delta N^2}^2$
	is a connected part of the forth moment. It clearly indicates the non-negligible $\mathcal{O}(1)$ NG.
	}
	\label{skew and kurt}
\end{figure}

On the other hand, the curvature perturbations produced around the critical point are naively thought to have NG as indicated by its non-perturbativity.
In figure~\ref{skew and kurt}, we show the skewness $S^{(3)}=\braket{\delta N^3}/\braket{\delta N^2}^{3/2}$ and kurtosis 
$S^{(4)}=\braket{\delta N^4}_c/\braket{\delta N^2}^2$ where $\braket{\delta N^4}_c=\braket{\delta N^4}-3\braket{\delta N^2}^2$ is a connected part
of the forth moment. 
These values vanish in a pure Gaussian case, so non-zero values of them directly indicate the NG of the curvature perturbations.
As we predicted, these plots show non-negligible $\mathcal{O}(1)$ NG.\footnote{However the NG is not as large as has been considered. 
Refs.~\cite{Lyth:2010zq,Bugaev:2011qt} concluded the curvature perturbations
have a negative chi-squared type distribution, namely $\zeta(\mathbf{x})=-(g^2(\mathbf{x})-\braket{g^2})$, 
where $g$ is a Gaussian field. But this distribution type gives $S^{(3)}=-2\sqrt{2}$ and
$S^{(4)}=12$ which are larger enough than our result. 
Therefore our calculations give less NG curvature perturbations than the simple chi-squared ansatz.}

These NG modifies the PDF of the curvature perturbations and then $\beta$ is given by,
\begin{align}\label{betaNG}
	\beta_\mathrm{NG}&\simeq\frac{2}{\sqrt{2\pi}}\int^\infty_\nu\dd\alpha\exp\left[\sum^4_{n=3}\frac{(-1)^n}{n!}S^{(n)}\pdif{{}^n}{\alpha^n}\right]\exp\left[-\frac{\alpha^2}{2}\right] \nonumber \\
	&\simeq\sqrt{\frac{2}{\pi}}\frac{1}{\nu}\exp\left[\sum^4_{n=3}\frac{\nu^2}{n!}S^{(n)}\right]\ee^{-\nu^2/2},
\end{align}
where $\nu$ is defined by $\nu=\zeta_c/\sigma_s$ and we used the high peak limit $\nu\gg1$ in the second line.
Though we truncated them here, it is possible to include the higher order terms than forth order (see also appendix K of ref.~\cite{Jeong}).
This modified probability is plotted on the right panel in figure~\ref{beta and betaNG}. 
Compared to the Gaussian case (left panel), it can be seen that the probability for small $\Pi^2$ is suppressed for $\zeta_c=1$, 
while the result for $\zeta_c=0.086$ hardly changes. 
As a result, the constraint for $\Pi^2$ in the $\zeta_c=1$ case is weakened to $\Pi^2\ltsim11$, 
which corresponds with $\braket{N}_\mathrm{water}\ltsim4$.
	
Let us briefly summarize the above results here. We calculated e-folds numerically 
in the stochastic formalism and checked that the mean e-folds for the waterfall phase and their variance depend almost only on
some specific parameter combination $\Pi^2=M^2\phi_c\mu_1/M_p^4$. 
However simultaneously it was found that the variance of the perturbations becomes large for mild-waterfall hybrid inflation. In fact,
if $\zeta_c=0.086$~\cite{Harada:2013epa}, there is no parameter region where PBHs are not overproduced. 
Only if the PBH mass given by eq.~(\ref{MPBH}) is lighter than $10^9\,\mathrm{g}$,
the constraint can be avoided because such PBHs are evaporated before big-bang nucleosynthesis (BBN).
On the other hand, if the threshold is as high as $\zeta_c=1$, the PBH overproduction roughly gives the parameter constraint 
as $\Pi^2\ltsim8$ (or 11 with NG corrections)
which means the waterfall phase can continue few e-folds. 
It is too short to produce PBHs massive enough to be DMs or seeds of SMBHs.
This is the main result of this paper.
  
Here note that the coarse-graining scale is the Hubble scale at the end of inflation, which is the smallest scale, and therefore $\beta$ shown in this section includes
all contributions of various mass PBHs. However, as we will see in the next section, the power spectrum has large and only one peak and therefore the main contribution for $\beta$
is given almost only by the PBHs whose mass corresponds with the peak scale. Hence the resulting $\beta$ for peak scale can be approximated well by that obtained in this section.

\section{Examples of power spectrum}\label{Example power spectrum}
In the previous section, we estimated the PBH abundances with use of the curvature perturbations coarse-grained on the Hubble scale at the end of inflation.
For them to be a good approximation, it is needed that the power spectrum has only one large peak. Moreover it is still unknown where the peak scale is.
To clarify them, in this section we show some examples of the power spectra calculated in the stochastic-$\delta N$ algorithm~\cite{Fujita:2013cna,Fujita:2014tja}.

Let us briefly review the algorithm at first. As mentioned repeatedly, in the stochastic formalism the dynamics in each Hubble patch is 
treated as an individual Brownian motion drifted by the potential force. The correlative information for two distant points is imprinted in the time 
when those two points are separated farther than the horizon scale. That is, while the scalar fields on two points evolves conjointly before the horizon exit,
they start moving individually due to the non-correlating noise after their distance is larger than the Hubble size.  
It means the perturbations of the e-folding numbers obtained for the paths branching from one field-phase-space point include
only the modes which exit the horizon between the branching time and the end of inflation. Thus the following relation 
is satisfied.
\begin{align}
	\braket{\delta N^2}=\int^{\log k_f}_{\log k_f-\braket{N}}\mathcal{P}_\zeta\,\dd\log k,
\end{align}
where $\braket{N}$ is the mean e-folding number from the branching point to the end of inflation.
Inversely the power spectrum can be obtained from changing the branching point slightly as,
\begin{align}
	\mathcal{P}_\zeta(k)=\left.\dif{}{\braket{N}}\braket{\delta N^2}\right|_{\braket{N}=\log(k_f/k)}.
\end{align} 
Here note that $k_f$ represents rather $\epsilon aH|_f$ than $aH|_f$.

While the branching point vev and the mean e-folds have a one-to-one correspondence in the single-field slow-roll case, 
there is only the uniform mean e-folds hypersurface on the field phase space in the multi-field case.
Therefore the branching points should be properly weighted by the realization probability, which is reproduced by making various sample paths.
Concretely, we use the following algorithm.
\begin{itemize}
\item[1.] 
Determine the initial field value from which the mean e-folds is about 60. It represents the field value of our observable universe at 
60 e-folds before the end of inflation.\footnote{The power spectrum obtained by the following procedure depends on this initial field value in
principle. This ambiguity is not only for our algorithm. The predictability of the inflation model generically reduces unless the inflatons' trajectory
converge well at least on the CMB scale. In our hybrid inflation case, the CMB scale mode exits the horizon in the well-converging valley phase,
and indeed the result is not affected by the initial condition so much.}

\item[2.]
Make one sample path by integrating the Langevin eqs.~(\ref{langevin}) and (\ref{calPphi}) from the initial field value. It shows the dynamics of some Hubble patch in our 
observable universe.

\item[3.]
Produce various realizations branching from some point on the produced sample path and calculate the e-folds for them to the final uniform density surface 
around the end of inflation. They give the mean and variance of the e-folds, referred as $\braket{N_1}$ and $\braket{\delta N_1^2}$ here.

\item[4.]
Repeat the procedure 3 with slightly different branching point on the same sample path to obtain another set of the mean and variance 
$\braket{N_2}$ and $\braket{\delta N_2^2}$. Then the power spectrum on the scale $k\simeq k_f\ee^{-(\braket{N_1}+\braket{N_2})/2}$
can be approximated by,
\begin{align}
	\mathcal{P}_\zeta(k)\simeq\frac{\braket{\delta N_1^2}-\braket{\delta N_2^2}}{\braket{N_1}-\braket{N_2}}.
\end{align}
This power spectrum is obtained from the paths branching from one sample path. Therefore this result is valid only in the region spatially near to 
that sample path. 

\item[5.]
To obtain the power spectrum valid over our universe, iterate the procedure 2--4 and average the obtained power spectra. This 
averaged one represents the true power spectrum obtained in our observable universe.

\end{itemize} 

\begin{figure}
	\centering
	\includegraphics[width=\hsize]{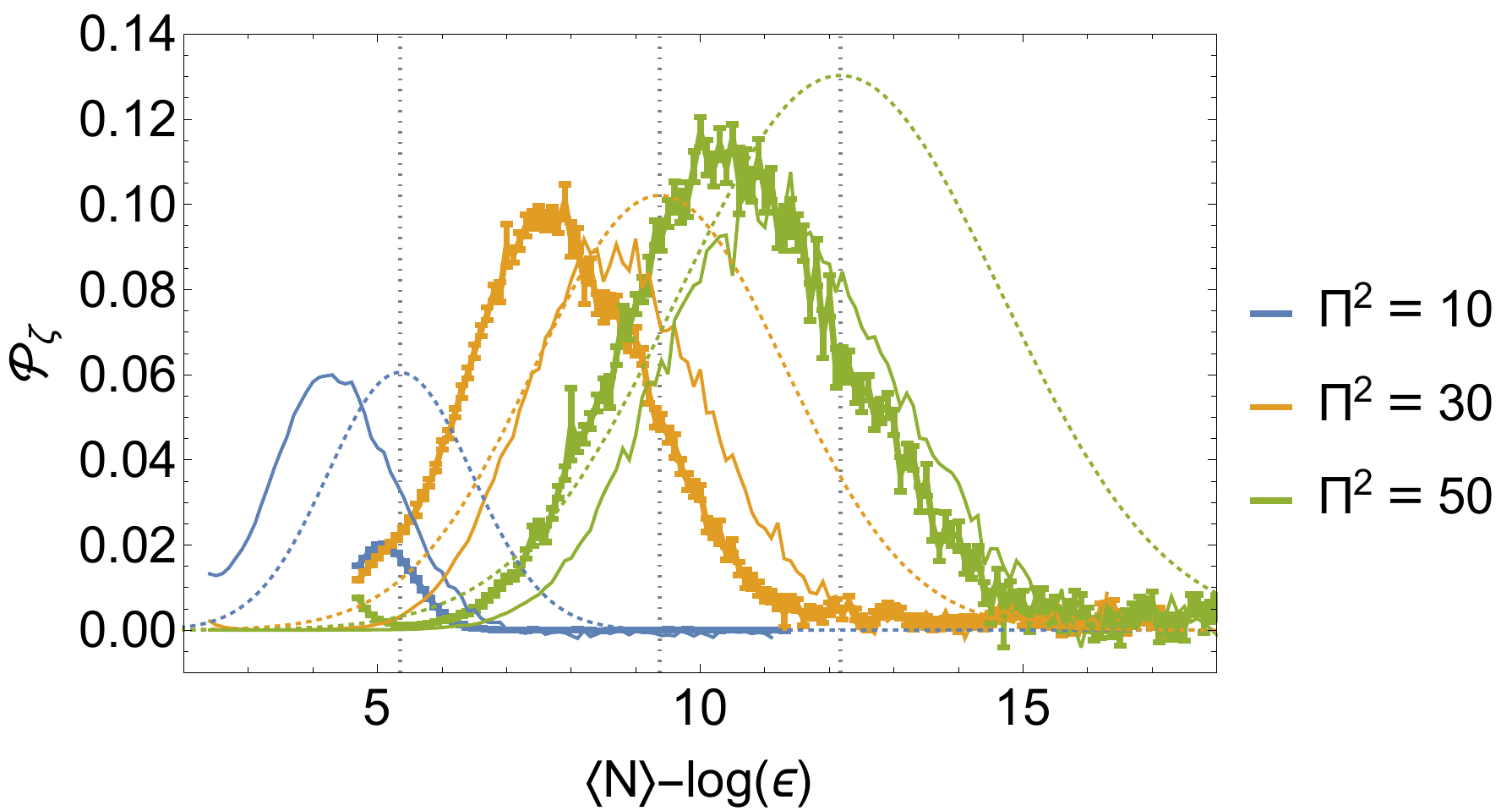}
	\caption{
	The power spectrum calculated in the stochastic-$\delta N$ algorithm~\cite{Fujita:2013cna,Fujita:2014tja}. 
	The thick lines with error bars represent the results for $\epsilon=0.01$, while the plane and dotted lines denote those for $\epsilon=0.1$ 
	and the CG's analytic approximations respectively.
	For $\epsilon=0.1$, we omit the error bars to avoid a busy figure.
	The color variation represents the difference of $\Pi^2$, but $M$ and $\phi_c$ are fixed to $0.1M_p$ and $0.1\sqrt{2}M_p$ respectively.
	Each power spectrum is averaged over 2500 sample paths and on each data point 1000 paths are made for each sample path.
	The error bars represent the standard errors.
	The horizontal axis shows the corresponding scale including $-\log\epsilon$ to cancel the scale shift due to the variation of $\epsilon$.
	From left to right, the vertical gray dotted lines represent the times when the paths pass the critical point for $\Pi^2=10,\,30,$ and $50$.
	It suggests that the power spectrum has a peak slightly after the critical point, which reflects that the $\psi$'s noise itself becomes slightly large 
	after the critical point due to its tachyonic mass.
	}
	\label{power_spectra}
\end{figure}

With use of the above algorithm, we calculate the power spectrum for $\Pi^2=10,\,30,$ and $50$, and the results are shown 
in figure~\ref{power_spectra}, compared to those for $\epsilon=0.1$ and the CG's analytic approximations.
While they are not so different for $\Pi^2=30$ or $50$, the peak for $\epsilon=0.01$ 
is much smaller than that for $\epsilon=0.1$ or the CG's one for $\Pi^2=10$.
This is simply because $-\log0.01\simeq4.61$ is larger than the peak scale $N_\mathrm{peak}\sim4$. 
Namely the peak scale is smaller than $(\epsilon aH|_f)^{-1}$ for $\epsilon=0.01$, 
which is the smallest scale the stochastic formalism can treat. 
Therefore the result for low $\Pi^2$ around 10 may be unreliable, but anyway the corresponding PBH mass 
$M_\mathrm{PBH}\sim\frac{M_p^2}{H_\mathrm{inf}}\ee^{2\times4}=3.0\times10^7\,\mathrm{g}\left(\frac{H_\mathrm{inf}}{10^9\,\mathrm{GeV}}\right)^{-1}$ is still too small
and our main conclusion that massive PBHs are overproduced in hybrid inflation is not changed.

\section{Conclusions}\label{Conclusions}
In this paper we study the possibility whether massive PBHs can be produced in mild-waterfall hybrid inflation
whose potential can be parametrized as eq.~(\ref{potential}).
As a recent related work, Clesse and Garcia-Bellido (CG)~\cite{Clesse:2015wea} 
estimated the curvature perturbations during the waterfall phase in the linear $\delta N$
formalism and found that the results depend almost only on the specific parameter combination $\Pi^2=M^2\phi_c\mu_1/M_p^4$ as
briefly reviewed in section~\ref{Aspects of hybrid inflation}.
However, since the waterfall field dynamics is dominated by the Hubble fluctuations around the critical point, 
one must go beyond the linear perturbation theory to calculate the curvature perturbations.
Accordingly we calculate the curvature perturbations without perturbative expansions with respect to the scalar fields, 
combining the stochastic and $\delta N$ formalism.

In section~\ref{Parameter search}, the variance of the curvature perturbations is calculated for various parameter values 
as shown in figure~\ref{N and dN2}, and it shows that indeed the curvature perturbations depend almost only on $\Pi^2$ though 
there are factor differences between our and CG's result. Subsequently we estimate the PBH abundance with use of this variance including
the non-Gaussian effect as plotted in figure~\ref{beta and betaNG}. The abundance is calculated both for the simple assumption of the 
threshold, namely $\zeta_c=1$, and the recent analytic work $\zeta_c\simeq0.086$~\cite{Harada:2013epa}. As a result, the specific parameter
combination $\Pi^2=M^2\phi_c\mu_1/M_p^4$ should be less than around 11 not to overproduce PBHs if $\zeta_c=1$, which means that 
the waterfall phase can continue only 4--5 e-folds at most. Also, if $\zeta_c\simeq0.086$, $\Pi^2$ should be smaller than $\sim1$.
In section~\ref{Example power spectrum}, we show some examples of the power spectra calculated in the stochastic-$\delta N$ formalism~\cite{Fujita:2013cna}.
They indicate that the power spectrum has one large peak slightly after the critical point.

In summary, the specific parameter combination $\Pi^2=M^2\phi_c\mu_1/M_p^4$ should be less than around 11 for $\zeta_c=1$
and then the waterfall phase cannot continue so long. Moreover in such cases the power spectrum shows a peak near the end of 
inflation $N_\mathrm{peak}\ltsim4$ and therefore the corresponding PBH mass scale is given by 
$M_\mathrm{PBH}\ltsim\frac{M_p^2}{H_\mathrm{inf}}\ee^{2\times4}=3.0\times10^7\,\mathrm{g}\left(\frac{H_\mathrm{inf}}{10^9\,\mathrm{GeV}}\right)^{-1}$.
It is much smaller than the desired one (comparable to $M_\odot\sim10^{33}\,\mathrm{g}$) because the corresponding scale does not benefit
by the exponential inflating during the waterfall phase so much.

As briefly mentioned in footnote~\ref{No of waterfall}, ref.~\cite{Halpern:2014mca} claimed that the power spectrum can be suppressed by the factor $\mathcal{N}$
with $\mathcal{N}$-waterfall fields. Therefore, if there are about $100$ waterfall fields and a continuous symmetry is imposed on them, one can obtain the desirable amplitude of the perturbations as
$\braket{\delta N^2}\sim0.01$ even for large $\Pi^2$. To verify such a dependence of the power spectrum on the number of the waterfall fields in a non-perturbative way,
we should apply the stochastic-$\delta N$ formalism for the multi-waterfall cases. It is beyond the scope of this paper and left for future works.

\acknowledgments
This work is supported by MEXT KAKENHI Grant Number 15H05889 (M. K.), JSPS KAKENHI Grant Number 25400248 (M. K.) 
and also by the World Premier International Research Center Initiative (WPI), MEXT, Japan. 
Y. T. is supported by a JSPS Research Fellowship for Young Scientists.



\end{document}